\documentclass[12pt,a4paper]{article}

\usepackage{amsmath,amsfonts,amssymb,bm}

\usepackage{graphicx} % do wczytywania rysunkow

\usepackage[hang,normalsize,bf]{subfigure} % rysuje kilka rysunkow
\begin{document}

\begin{center}

{\Huge \bf
Gluon transverse momenta\\ and\\
charm quark-antiquark \\ pair production \\
in $p-\bar p$ collisions at Tevatron
}

\vspace {0.6cm}

{\large M. {\L}uszczak $^{2}$ and A. Szczurek $^{1,2}$}

\vspace {0.2cm}

$^{1}$ {\em Institute of Nuclear Physics PAN\\
PL-31-342 Cracow, Poland\\}
$^{2}$ {\em University of Rzesz\'ow\\
PL-35-959 Rzesz\'ow, Poland\\}

\end{center}

\begin{abstract}
We discuss and compare different approaches to include gluon 
transverse momenta for heavy quark-antiquark pair production.
The correlations in azimuthal angle and in heavy quark, heavy
antiquark transverse momenta are studied in detail.
The results are illustrated with the help
of different unintegrated gluon distribution functons (UGDF)
from the literature. We compare results obtained with
on-shell and off-shell matrix elements and kinematics
and quantify where these effects are negligible and where they are
essential.
We concentrate on the region of asymmetric
transverse momenta of charm quark and charm antiquark.
Most of UGDFs lead in this corner of the phase space to almost full
decorellation in azimuthal angle.
We propose correlation observables to be best suited
in order to test the existing models of UGDFs.
\end{abstract}

PACS: 12.38.-t, 12.38.Cy, 14.65.Dw 

%---------------------
\section{Introduction}
%---------------------

The charm-anticharm production in photo- and hadroproduction
is known as one of the crucial tests of conventional 
gluon distributions within a standard factorization approach.
At high energies one tests gluon distributions at low values
of longitudinal momentum fraction. At leading order
of collinear approach the heavy quark and heavy antiquark
are produced back to back i.e. the azimuthal correlation function
is proportional to $\delta(\phi-\pi)$ and the distribution in
$p_{1,t}$ (heavy quark transverse momentum) and $p_{2,t}$
(heavy antiquark transverse momentum) is proportional to
$\delta(p_{1,t}-p_{2,t})$.
In collinear approach this changes only in next-to-leading order.
In unintegrated gluon distribution approach the azimuthal angle
and $p_{t}$ decorrelations are obtained already in leading order
of perturbative expansion.

Different models of UGDF were presented in the literature.
Recently we have tested some of the models in photoproduction
\cite{LS04}. Here we shall discuss in detail the case of hadroproduction.

It was realized recently that the understanding
of transverse momenta can be a clue to understand many
high energy reactions.
In some reactions like meson hadroproduction, prompt photon
and Drell-Yan production or single spin asymmetry (SSA)
the transverse momenta are not obtained dynamically but are included
``by hand'' with Gaussian smearing
(see e.g. \cite{Gaussian_smearing,AM04}).
The parameter of the smearing is assumed to be independent of
longitudinal momentum fraction and is usually arbitrarily chosen or
adjusted to a selected set of data.

In some of high-energy reactions that involve gluons the unintegrated
gluon distributions are modelled or are obtained as a solution
of some (usually simplified) QCD evolution equations. In particular,
the charm-anticharm hadroproduction is considered as one of
the flag processes for the so-called $k_{t}$-factorization approach
\cite{CCH91,CE91,BE01}.
There are already a few phenomenological trials to calculate
the charm or bottom production within the $k_{t}$-factorization
approach \cite{RSS,BS00,LSZ02}. The results of these trials differ
considerably due to completely different UGDFs used.

In the present paper we shall compare different approaches
how to include transverse momenta proposed
in the literature for the charm-anticharm pair production
at the Tevatron energy W = 1.96 GeV both for the single
charm/anticharm distributions as well as for a few charm-anticharm
correlation observables.

%------------------
\section{Formalism}
%------------------

Let us consider the reaction $h_1 + h_2 \to Q + \bar Q + X$,
where $Q$ and $\bar Q$ are heavy quark and heavy antiquark,
respectively.
In the leading-order (LO) approximation within collinear approach
the triple-differential cross section in rapidity of $Q$ ($y_1$),
in rapidity of $\bar Q$ ($y_2$) and transverse momentum of
one of them ($p_t$) can be written as
\begin{equation}
\frac{d \sigma}{d y_1 d y_2 d^2p_t} = \frac{1}{16 \pi^2 {\hat s}^2}
\sum_{i,j} x_1 p_i(x_1,\mu^2) \; x_2 p_j(x_2,\mu^2) \;
\overline{|{\cal M}_{ij}|^2} \; .
\label{LO_collinear}
\end{equation}
Above $p_i(x_1,\mu^2)$ and $p_j(x_2,\mu^2)$ are familiar
(integrated) parton distributions in hadron $h_1$ and $h_2$, respectively.
There are two types of LO $2 \to 2$ subprocesses which enter
Eq.(\ref{LO_collinear}): $gg \to Q \bar Q$ and $q \bar q \to Q \bar Q$.
The first mechanism dominates at large energies and the second near the
threshold. The parton distributions are evaluated at:
$x_1 = \frac{m_t}{\sqrt{s}}\left( \exp( y_1) + \exp( y_2) \right)$,
$x_2 = \frac{m_t}{\sqrt{s}}\left( \exp(-y_1) + \exp(-y_2) \right)$,
where $m_t = \sqrt{p_t^2 + m_Q^2}$.
The formulae for matrix element squared averaged over initial
and summed over final spin polarizations can be found e.g. in
Ref.\cite{BP_book}.

If one allows for transverse momenta of the initial partons,
the sum of transverse momenta of the final $Q$ and $\bar Q$ no longer
cancels.
Formula (\ref{LO_collinear}) can be easily generalized if
one allows for the initial parton transverse momenta. Then
\begin{eqnarray}
\frac{d \sigma}{d y_1 d y_2 d^2p_{1,t} d^2p_{2,t}} = \sum_{i,j} \;
\int \frac{d^2 \kappa_{1,t}}{\pi} \frac{d^2 \kappa_{2,t}}{\pi}
\frac{1}{16 \pi^2 (x_1 x_2 s)^2} \; \overline{ | {\cal M}_{ij} |^2}
\nonumber \\  
\delta^{2} \left( \vec{\kappa}_{1,t} + \vec{\kappa}_{2,t} 
                 - \vec{p}_{1,t} - \vec{p}_{2,t} \right) \;
{\cal F}_i(x_1,\kappa_{1,t}^2) \; {\cal F}_j(x_2,\kappa_{2,t}^2) \; ,
\label{LO_kt-factorization}    
\end{eqnarray}
where now ${\cal F}_i(x_1,\kappa_{1,t}^2)$ and ${\cal F}_j(x_2,\kappa_{2,t}^2)$
are so-called unintegrated parton distributions
\footnote{In this paper we shall use parallel two different conventions
of unintegrated gluon distributions:\\
(a) $\int_0^{\mu^2} {\cal F}(x,\kappa^2) d \kappa^2 \sim x g(x,\mu^2)$,\\
(b) $\int_0^{\mu^2} f(x,\kappa^2) \frac{d \kappa^2}{\kappa^2}
 \sim x g(x,\mu^2)$.
}.
The extra integration is over transverse momenta of the initial
partons.
The two extra factors $1/\pi$ attached to the integration over
$d^2 \kappa_{1,t}$ and $d^2 \kappa_{2,t}$ instead over
$d \kappa_{1,t}^2$ and $d \kappa_{2,t}^2$ as in the conventional
relation between unintegrated ($\cal F$) and integrated ($g$) parton
distributions. 
The two-dimensional Dirac delta function assures momentum conservation.
Now the unintegrated parton distributions must be evaluated at:
$x_1 = \frac{m_{1,t}}{\sqrt{s}}\left( \exp( y_1) + \exp( y_2) \right)$,
$x_2 = \frac{m_{2,t}}{\sqrt{s}}\left( \exp(-y_1) + \exp(-y_2)
\right)$, where $m_{i,t} = \sqrt{p_{i,t}^2 + m_Q^2}$.
In general, the matrix element must be calculated for initial
off-shell partons. The corresponding formulae for initial gluons
were calculated in \cite{CCH91,CE91} (see also \cite{BE01}).
In the present paper for illustration we shall compare results
obtained for both on-shell and off-shell matrix elements.
It is easy to check that in the limit
$\kappa_1^2 \to 0$, $\kappa_2^2 \to 0$
\begin{equation}
\begin{split}
&|{\cal M}_{gg \to Q\bar Q}^{off-shell}|^2 \to
 |{\cal M}_{gg \to Q\bar Q}^{on-shell}|^2 =
2 \pi^2 \alpha_s^2(...)  \\
&\left[
\frac{6(m_Q^2 - \hat t)(m_Q^2 - \hat u)}{\hat s^2} 
- \frac{m_Q^2(\hat s - 4 m_Q^2)}{3(m_Q^2-\hat t)(m_Q^2-\hat u)}\right. \\
&+\frac{4}{3} \cdot 
\frac{(m_Q^2-\hat t)(m_Q^2-\hat u)-2 m_Q^2(m_Q^2+\hat t)}
{(m_Q^2 - \hat t)^2}
+\frac{4}{3} \cdot 
\frac{(m_Q^2-\hat u)(m_Q^2-\hat t)-2 m_Q^2(m_Q^2+\hat u)}
{(m_Q^2 - \hat u)^2}    \\
&\left.
-3 \cdot \frac{(m_Q^2-\hat t)(m_Q^2-\hat u)+m_Q^2(\hat u - \hat t)}
{\hat s(m_Q^2 - \hat t)}
-3 \cdot \frac{(m_Q^2-\hat u)(m_Q^2-\hat t)+m_Q^2(\hat t - \hat u)}
{\hat s(m_Q^2 - \hat u)}
\right]  \; ,
\end{split}
\label{on-shell-ME2}
\end{equation}
where $\hat t, \hat u, \hat s$ are Mendelstam variables
for the partonic subprocess.

Introducing new variables:
\begin{eqnarray}
\vec{Q}_t = \vec{\kappa}_{1,t} + \vec{\kappa}_{2,t} \; , \nonumber \\
\vec{q}_t = \vec{\kappa}_{1,t} - \vec{\kappa}_{2,t} \; 
\label{new_variables}
\end{eqnarray}
we can write:
\begin{eqnarray}
\frac{d \sigma_{ij}}{d y_1 d y_2 d^2p_{1,t} d^2p_{2,t}} =
\int d^2 q_t \; \frac{1}{4 \pi^2}
\frac{1}{16 \pi^2 (x_1 x_2 s)^2} \; \overline{ | {\cal M}_{ij} |^2}
\nonumber \\  
{\cal F}_i(x_1,\kappa_{1,t}^2) \; {\cal F}_j(x_2,\kappa_{2,t}^2) \; .
\label{LO_kt-factorization2}    
\end{eqnarray}
This formula is very useful to study correlations between
the produced heavy quark $Q$ and heavy antiquark $\bar Q$.

For example
\begin{eqnarray}
\frac{d \sigma_{ij}}{d p_{1,t} d p_{2,t}} &=&
\int d \phi_1 d \phi_2 \; p_{1,t} p_{2,t} \int dy_1 d y_2
\int d^2 q_t \; \frac{1}{4 \pi^2}
\frac{1}{16 \pi^2 (x_1 x_2 s)^2} \; \overline{ | {\cal M}_{ij} |^2}
\nonumber \\  
&&{\cal F}_i(x_1,\kappa_{1,t}^2) \; {\cal F}_j(x_2,\kappa_{2,t}^2) 
\nonumber \\
&=& 4 \pi \; \frac{1}{2} \; \frac{1}{2} \;
 \int d \phi_{-} \; p_{1,t} p_{2,t} \int dy_1 d y_2
\int d^2 q_t \; \frac{1}{4 \pi^2}
\frac{1}{16 \pi^2 (x_1 x_2 s)^2} \; \overline{ | {\cal M}_{ij} |^2}
\nonumber \\  
&&{\cal F}_i(x_1,\kappa_{1,t}^2) \; {\cal F}_j(x_2,\kappa_{2,t}^2)  \; .
\label{p1t_p2t_map}
\end{eqnarray}
In the last equation we have introduced $\phi_{-} \equiv \phi_1 -
\phi_2$, where $\phi_{-} \in$ (-2$\pi$, 2$\pi$).
The factor 4 $\pi$ comes from the integration over
$\phi_{+} \equiv \phi_1 + \phi_2$. The first factor 1/2 comes from
the jacobian transformation while the second factor 1/2
takes into account an extra extension of the domain
when using $\phi_{+}$ and $\phi_{-}$ instead of $\phi_{1}$
and $\phi_{2}$.

At the Tevatron energy the contribution of the $gg \to Q \bar Q$
subrocess is more than order of magnitude larger than its
counterpart for the $q \bar q \to Q \bar Q$ subprocess.
Therefore in the following we shall take into account
only gluon-gluon fusion process i.e. i=0 and j=0.
The quark-antiquark mechanism
is important only at small center-of-mass energies.

In LO collinear approach
\begin{equation}
\frac{d \sigma}{d p_{1,t} d p_{2,t}} \propto \delta(p_{1,t} - p_{2,t})
\; .
\label{collinear_LO_p1t_p2t_map}
\end{equation}
A deviation from this relation is therefore a measure of the initial gluon
(parton) transverse momenta.

Purely perturbative \footnote{when both UGDFs are generated perturbatively}
$k_t$-factorization formalism to $h_1 h_2 \to Q \bar Q$ applies if
$\kappa_{1,t}^2, \kappa_{2,t}^2 > \kappa_0^2$.
The choice of $\kappa_0^2$ is to large extent arbitrary.
In Refs.\cite{RSS} a rather large $\kappa_0^2$ was chosen
and the space $\kappa_{1,t}^2 \times \kappa_{2,t}^2$ was
subdivided into four disjoint regions. For example
the contribution when both $\kappa_{1,t}^2$ and $\kappa_{2,t}^2$
are small was replaced by the leading-order collinear cross section.
Such an approach by construction assures that
$\sigma_{Q \bar Q}^{tot} > \sigma_{Q \bar Q}^{tot}$(collinear LO). \\
It is rather obvious that the total cross section strongly depends
on the choice of $\kappa_0^2$. Our philosophy here is different.
Many models of UGDF in the literature treat the soft region explicitly.
Therefore we use the $k_t$-factorization formula everywhere on
the $\kappa_{1,t}^2 \times \kappa_{2,t}^2$ plane. For
perturbatively generated UGDFs, like KMR \cite{KMR} for instance,
we shall extrapolate the perturbative behaviour into the soft region.
Thus our approach guarantees a smooth behaviour on
the $\kappa_{1,t}^2 \times \kappa_{2,t}^2$ plane and reduces
arbitrariness in dividing the two-dimensional gluon-transverse-momentum space.

%-----------------------------------------
\section{Unintegrated gluon distributions}
%-----------------------------------------

Obtaining UGDFs from underlying QCD is not an easy task.
The main reason is the difficulty in separating out the perturbative
from nonperturbative domains. The nonperturbative domain is
difficult {\it per se}. Even in perturbative region different schemes of
resummation have been proposed.
In this section we describe briefly some representative UGDFs
from the literature used in the present paper to calculate
charm-anticharm production.

%-----------------------------------------
\subsection{Kwieci\'nski gluon distribution}
%-----------------------------------------

Kwieci\'nski has shown that the evolution equations
for unintegrated parton distributions takes a particularly
simple form in the variable conjugated to the parton transverse momentum.
The two possible representations are interrelated via Fourier-Bessel
transform
\begin{equation}
  \begin{split}
    &{{\cal F}_k(x,\kappa_t^2,\mu_F^2) =
    \int_{0}^{\infty} db \;  b J_0(\kappa_t b)
    {\tilde {\cal F}}_k(x,b,\mu_F^2)} \; ,
    \\
    &{{\tilde {\cal F}}_k(x,b,\mu_F^2) =
    \int_{0}^{\infty} d \kappa_t \;  \kappa_t J_0(\kappa_t b)
    {\cal F}_k(x,\kappa_t^2,\mu_F^2)} \; .
  \end{split}
\label{Fourier}
\end{equation}
The index k above numerates either gluons (k=0), quarks (k$>$ 0) or
antiquarks (k$<$ 0).

In the impact-parameter space the Kwieci\'nski equation
takes a rather simple (diagonal in b) form
\cite{CCFM_b1,CCFM_b2,GKB03}
(see also \cite{CS05}).

The perturbative solutions ${\tilde{\cal F}}_q^{pert}(x,b,\mu_F^2)$ do not
include nonperturbative effects such as, for instance,
intrinsic transverse momenta of partons in colliding hadrons.
One of the reasons is e.g. internal motion of constituents of the proton. 
In order to include such effects we modify the perturbative
solution $\tilde{{\cal F}}_g^{pert}(x,b,\mu_F^2)$
and write the modified parton distributions
$\tilde{\cal {F}}_g(x,b,\mu_F^2)$ in the simple factorized form
\begin{equation}
\tilde{{\cal F}}_g(x,b,\mu_F^2) = \tilde{{\cal F}}_g^{pert}(x,b,\mu_F^2)
 \cdot F_g^{np}(b) \; .
\label{modified_uPDFs}
\end{equation}
In the present study we shall use the following functional
form for the nonperturbative form factor
\begin{equation}
F_g^{np}(b) = F^{np}(b) = \exp\left(- \frac{b^2}{4 b_0^2}\right) \; .
%\text{or} \; \exp \left( - \frac{b}{b_e} \right)
\label{formfactor}
\end{equation}
In Eq.(\ref{formfactor}) $b_0$ is the only free parameter.

In the following we use leading-order integrated parton distributions
from Ref.\cite{GRV98} as the initial condition for QCD evolution.
The set of integro-differential equations in b-space
is solved by the method based on the discretisation made with
the help of the Chebyshev polynomials (see e.g. \cite{GKB03}).
Then the unintegrated parton distributions $\tilde{{\cal F}}_g(x,b,\mu_F^2)$
are put on a grid in $x$, $b$ and $\mu^2$ and the grid was used in practical
applications for Chebyshev interpolation. 
In our practical application we need rather gluon distributions
in momentum space. The latter are obtained via Fourier transform
(\ref{Fourier}) from $\tilde{{\cal F}}_g(x,b,\mu_F^2)$.

%-----------------------------------------
\subsection{Kimber-Martin-Ryskin distribution}
%-----------------------------------------

%The unintegrated gluon distribution can be obtained even
%when the integrated gluon distribution fulfils standard
%DGLAP evolution equation.
%At very small~$x$
%
%\begin{equation}
%{\cal F}(x,\kappa^2) = \frac{\partial}{\partial Q^2}
%\left[ x g(x,Q^2) \right] |_{Q^2 = \kappa^2} \; .
%\label{derivative}
%\end{equation}
%
%This prescription breaks at larger values of $x$ when
%the derivative of the gluon distribution becomes negative.
%This may be somewhat improved by introducing a Sudakov form factor
%$T_g(\kappa^2,\mu^2)$.
%Then the unintegrated gluon distribution reads \cite{KMR}:
%
%\begin{equation}
%{\cal F}(x,\kappa^2,\mu^2) =
%\frac{\partial}{\partial Q^2}
%\left[
%T(Q^2,\mu^2) x g(x,Q^2) \right] |_{Q^2 = \kappa^2}
% \; .
%\label{KMR}
%\end{equation}

Resumming virtual contributions to DGLAP equation,
the unintegrated parton distributions can be written as \cite{KMR}
\begin{equation}
f_a(x,\kappa^2,\mu_F^2) = T_a(\kappa^2,\mu_F^2)
\cdot \frac{\alpha_s(\kappa^2)}{2 \pi}
\sum_{a'} \int_x^{1-\delta} P_{aa'}(z) \left(\frac{x}{z} \right)
a'\left(\frac{x}{z},\kappa^2\right) dz \; .
\label{KMR_master} 
\end{equation}
Specializing to the gluon distribution the Sudakov form factor
reads as
\begin{equation}
T_g(\kappa^2,\mu_F^2) =
\exp\left(       
-\int_{\kappa^2}^{\mu_F^2} \frac{d p^2}{p^2} \frac{\alpha_s(p^2)}{2 \pi}
\int_{0}^{1-\delta} dz \left[z P_{gg}(z)+\sum_q P_{qg}(z)     \right]
\right) \; .
\label{Sudakov}
\end{equation}
The Sudakov form factor introduces a dependence on a second scale
$\mu_F^2$.
Different prescriptions for $\delta$ have been used in the literature.
The most popular choice $\delta = \mu_F/(\mu_F+\kappa)$
corresponds to the angular ordering in the gluon emission.

It is reasonable to assume that the unintegrated gluon density
given by (\ref{KMR_master}) starts only
at $\kappa_t^2 > \kappa_{t0}^2$ \cite{KMS97}, i.e. in the perturbative
domain.
At lower $\kappa_t^2$ an extrapolation is needed.
A use of
the GRV integrated gluon distribution \cite{GRV95,GRV98}
in (\ref{KMR_master}) seems more adequate than any other gluon distribution
because it allows to go down to small gluon transverse momenta.
Following Ref.\cite{szczurek03} $\kappa_{t0}^2$ = 0.5 GeV$^2$ is taken
as the lowest value where the unintegrated gluon distribution is
calculated from Eq.(\ref{KMR_master}). Below it is assumed
\begin{equation}
{\cal F}(x,\kappa^2) = f(x,\kappa^2)/\kappa^2 =
f(x,\kappa_0^2)/\kappa_0^2 \; .
\label{low_kappa}
\end{equation}
In general, the quantity $\kappa_{t0}^2$ can be treated as a free
nonperturbative parameter.

The choice of $\mu_F^2$ in our case of charmed quark-antiquark production
is not completely obvious.
In the present analysis, because we limit to not too large
quark/antiquark transverse momenta, $\mu_F^2 = m_c^2, 4 m_c^2$
is taken for simplicity. This allows to prepare a two-dimensional grid
${\cal F}_{KMR}(x_i,\kappa_j^2)$
to be used for further interpolation in the $k_t$-factorization
formula which accelerates the multi-dimensional integrations.

If $T_g$ in Eq.(\ref{KMR_master}) is ignored we shall denote
the corresponding gluon distribution as $f_{DGLAP}$ or
${\cal F}_{DGLAP}$ and call it DGLAP gluon distribution for brevity.

%-----------------------------------
\subsection{BFKL gluon distribution}
%-----------------------------------

At very low $x$ the unintegrated gluon distributions are believed
to fulfil BFKL equation \cite{BFKL}.
After some simplifications \cite{AKMS94} the BFKL equation reads
\begin{equation}
-x \frac{\partial f(x, q_t^2)}{\partial x} =
\frac{\alpha_s N_c}{\pi} q_t^2
\int_0^{\infty} \frac{dq_{1t}^2}{q_{1t}^2}
\left[                                 
\frac{f(x,q_{1t}^2) - f(x,q_t^2)}{|q_t^2 - q_{1t}^2|}
+ \frac{f(x,q_t^2)}{\sqrt{q_t^4+4 q_{1t}^4}}
\right] \; .
\label{BFKL_equation}
\end{equation}
The homogeneous BFKL equation can be solved numerically \cite{AKMS94}.
Here in the practical applications we shall use a simple
parametrization for the solution \cite{ELR96}
\begin{equation}
f(x,\kappa_t^2) = \frac{C}{x^{\lambda}}
\left(\frac{\kappa_t^2}{q_0^2}\right)^{1/2}
\frac{{\tilde \phi}_0}{\sqrt{ 2 \pi \lambda''\ln(1/x)}}
\exp\left[ - \frac{\ln^2(\kappa_t^2/{\bar q}^2)}{r2\lambda ''\ln(1/x)}
\right]    \; .
\label{ELR_parametrization}
\end{equation}
In the above expression $\lambda = 4 {\bar \alpha}_s \ln2$,
$\lambda''$ = 28 ${ \bar \alpha}_s \zeta(3)$, ${\bar \alpha}_s
= 3 \alpha_s/\pi, \zeta(3)$ = 1.202. The remaining parameters were
adjusted in \cite{ELR96} to reproduce with a satisfactory accuracy
the gluon distribution which was obtained in \cite{AKMS94}
as the numerical solution of the BFKL equation. 
It was found that ${\bar q} = q_0$ = 1, $C {\bar \phi_0}$ = 1.19
and $r$ = 0.15 \cite{ELR96}.

%------------------------------------------------------
\subsection{Golec-Biernat--W\"usthoff gluon distribution}
%------------------------------------------------------

Another parametrization of gluon distribution in the proton
can be obtained based on the Golec-Biernat--W\"usthoff
parametrization of the dipole-nucleon cross section with
parameters fitted to the HERA data.
The dipole-nucleon cross section can be transformed to corresponding
unintegrated gluon distribution.
The resulting gluon distribution reads \cite{GBW_glue}:
\begin{equation}
\alpha_s {\cal F}(x,\kappa_t^2) =
\frac{3 \sigma_0}{4 \pi^2} R_0^2(x) \kappa_t^2 \exp(-R_0^2(x)
\kappa_t^2) \; ,
\label{GBW_glue}
\end{equation}
where
\begin{equation}
R_0(x) = \left( \frac{x}{x_0} \right)^{\lambda/2}
\frac{1}{GeV} \; .
\end{equation}
From their fit to the data: $\sigma_0$ = 29.12 mb,
$x_0$ = 0.41 $\cdot$ 10$^{-4}$,
$\lambda$ = 0.277.
In order to determine the gluon distribution 
we take somewhat arbitrarily $\alpha_s$ = 0.2.

%--------------------------------------------------
\subsection{Gluon distribution a la Kharzeev-Levin}
%--------------------------------------------------

Another parametrization, also based on the idea of gluon
saturation, was proposed in \cite{KL01}.
In contrast to the GBW approach \cite{GBW_glue}, where
the dipole-nucleon cross section is parametrized,
in the Karzeev-Levin approach it is the unintegrated gluon distribution
which is parametrized.
In the following we shall consider the most simplified
functional form:
\begin{eqnarray}
{\cal F}(x,\kappa^2) =
\begin{cases}
 f_0                              & \text{if} \; \kappa^2 < Q_s^2 , \\
 f_0 \cdot \frac{Q_s^2}{\kappa^2} & \text{if} \; \kappa^2 > Q_s^2 .
\end{cases}
\label{KL_glue}
\end{eqnarray}
The saturation momentum $Q_s$ is parametrized exactly as in the GBW
model $Q_s^2(x) =$ 1 GeV$^2$ $\cdot \left( \frac{x_0}{x}
\right)^{\lambda}$. 

The normalization constant $f_0$ was adjusted in \cite{szczurek03}
to roughly describe the HERA data.

%-----------------------------------
\subsection{Naive Gaussian smearing}
%-----------------------------------

For better understanding we shall
compare our results with the results obtained with
a simple unintegrated gluon distribution:
\begin{equation}
{\cal F}_{naive}(x,\kappa^2,\mu_F^2) = x g^{coll}(x,\mu_F^2)
\cdot f_{Gauss}(\kappa^2) \; ,
\label{naive_UGDF}
\end{equation}
where $g^{coll}(x,\mu_F^2)$ is a standard collinear (integrated)
gluon distribution and $f_{Gauss}(\kappa^2)$ is
a Gaussian two-dimensional function:
\begin{equation}
f_{Gauss}(\kappa^2) = \frac{1}{2 \pi \sigma_0^2}
\exp \left( -\kappa_t^2 / 2 \sigma_0^2 \right) / \pi \; .
\label{Gaussian}
\end{equation}
Such a phenomenological procedure is often used to improve
collinear calculations for small transverse momenta
\cite{Gaussian_smearing}.
The UGDF defined by Eq.(\ref{naive_UGDF}) and (\ref{Gaussian})
is normalized such that:
\begin{equation}
\int {\cal F}_{naive}(x,\kappa^2,\mu_F^2) \; d \kappa^2 = x
g^{coll}(x,\mu_F^2) \; .
\label{naive_normalization}
\end{equation}
%

%----------------
\section{Results}
%----------------

In the present analysis we shall discuss both inclusive spectra
of charm quarks/antiquarks as well as correlations between 
quark and antiquark. In order to obtain the integrated cross
section a 7-dimensional integration has to be performed.
We have carefully explored the 7-dimensional phase space in order
to optimize the integration. If not otherwise stated we take
the whole range of quark/antiquark rapidities:
-6 $< y_1 <$ 6 and -6 $< y_2 <$ 6.

\begin{figure}[!thb] % Figure 1
\begin{center}
\includegraphics[width=6cm]{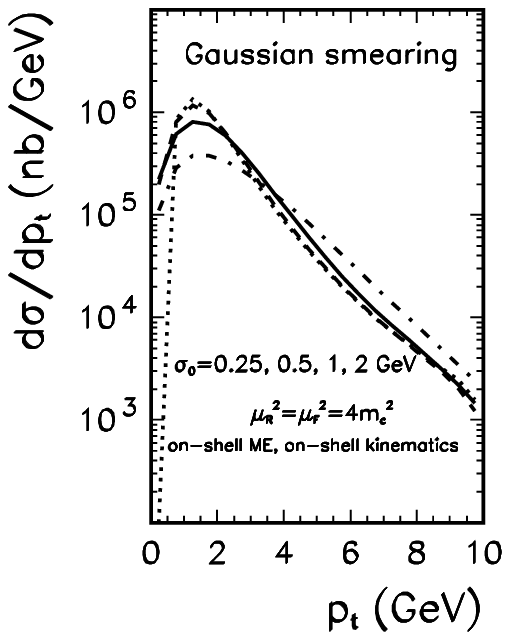}
\includegraphics[width=6cm]{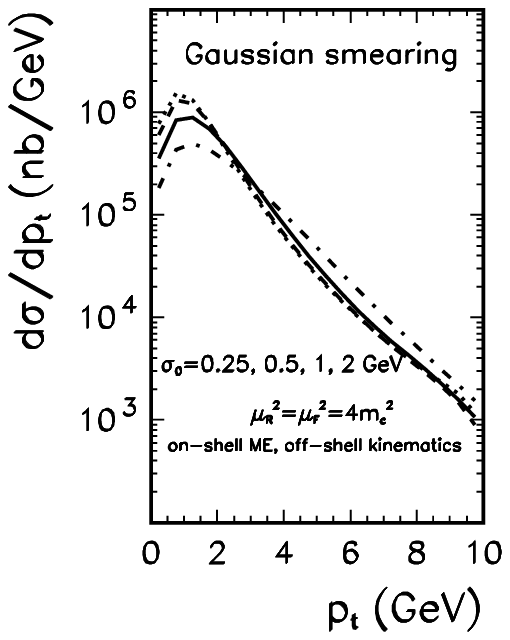}
\caption[*]{
Inclusive $d\sigma/d p_t$ for charm/anticharm production at W = 1.96 TeV in
the approach of naive Gaussian smearing of gluon transverse momenta.
The calculation with different values of
$\sigma_0 =$ 0.25 (dotted), 0.50 (dashed), 1.0 (solid), 2.0
(dash-dotted) GeV are shown.
The results with on-shell kinematics are shown in panel
(a) and results with off-shell kinematics in panel (b).
In this calculation both factorization and renormalization scales
were fixed for 4 $m_c^2$.
\label{fig:dsig_dpt_gauss}
}
\end{center}
\end{figure}

Let us start from single particle spectra of charm quarks.
In Fig.\ref{fig:dsig_dpt_gauss} we demonstrate the effect of
gluon transverse momentum Gaussian smearing on final quark transverse
momentum distribution for W = 1.96 TeV, i.e. at the present
Tevatron energy.
The smearing of primordial gluon distributions causses only
a mild broadening of the charm quark-antiquark spectra.
There is almost no difference if off-shell kinematics is used
instead of on-shell one that is usually used in the case
of the ad hoc Gaussian smearing.

\begin{figure}[!thb] % Figure 2
\begin{center}
\includegraphics[width=6.0cm]{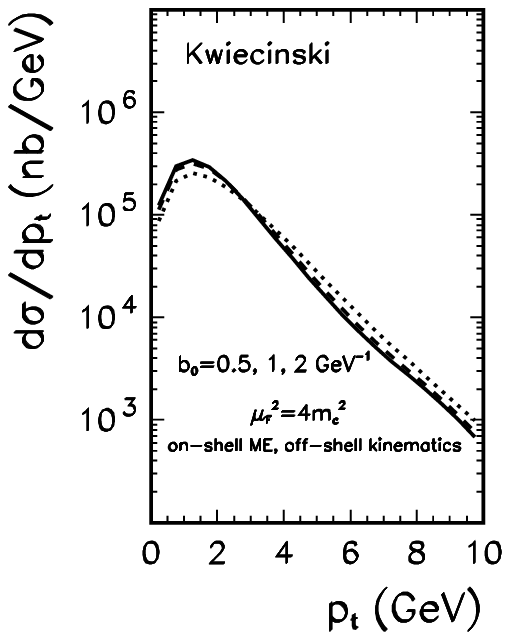}
\includegraphics[width=6.0cm]{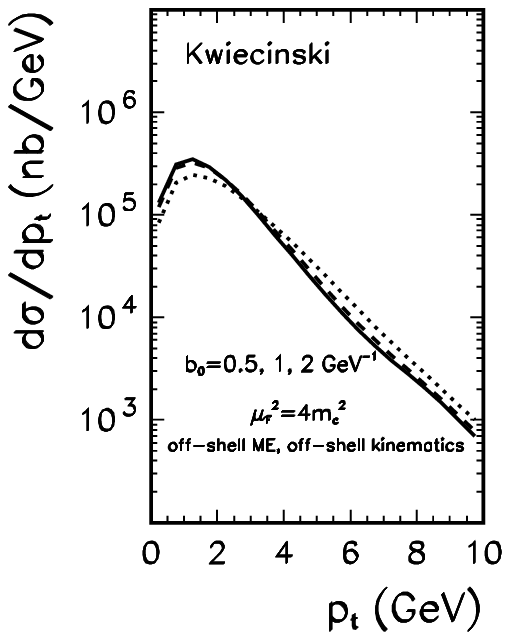}
\caption[*]{
Inclusive $d\sigma/d p_t$ for charm/anticharm production at W = 1.96 TeV for
the Kwieci\'nski unintegrated gluon distributions for
$b_0 =$ 0.5, 1.0, 2.0 GeV$^{-1}$ and for $\mu_F^2 = 4 m_c^2$.
The results with on-shell matrix element is shown in the left panel
and results with off-shell matrix element in the right panel.
\label{fig:dsig_dpt_kwiec_b0}
}
\end{center}
\end{figure}

\begin{figure}[!thb] % Figure 3
\begin{center}
\includegraphics[width=8.0cm]{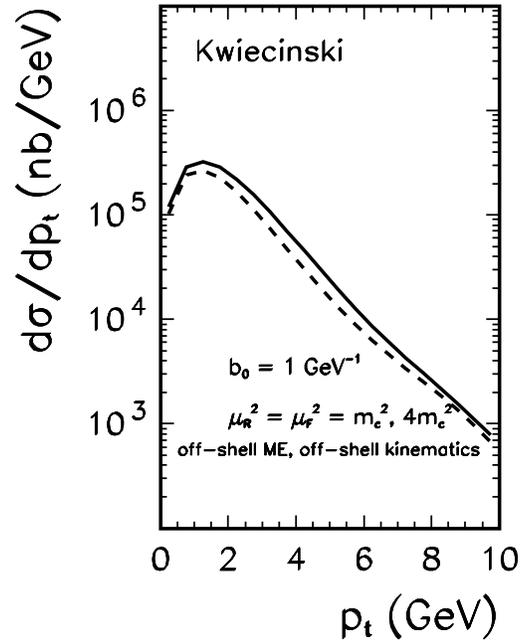}
\caption[*]{
Inclusive $d\sigma/d p_t$ for charm/anticharm production at W = 1.96 TeV
for the Kwieci\'nski UGDF for different factorization scales
$\mu_F^2$ = $m_c^2$ (dashed), 4 $m_c^2$ (solid).
In this calculation $b_0 =$ 1.0 GeV$^{-1}$.
\label{fig:dsig_dpt_kwiec_mu2}
}
\end{center}
\end{figure}

In Fig.\ref{fig:dsig_dpt_kwiec_b0} we show a similar effect
for the Kwieci\'nski UGDF for fixed factorization scale
$\mu_F^2 = 4 m_c^2$. There is rather weak dependence on
the value of $b_0$ used for the nonperturbative form factor.
The dependence on factorization scale is shown in
Fig.\ref{fig:dsig_dpt_kwiec_mu2}. The resulting effect is rather mild.

\begin{figure}[!thb] % Figure 4
\begin{center}
\includegraphics[width=8.0cm]{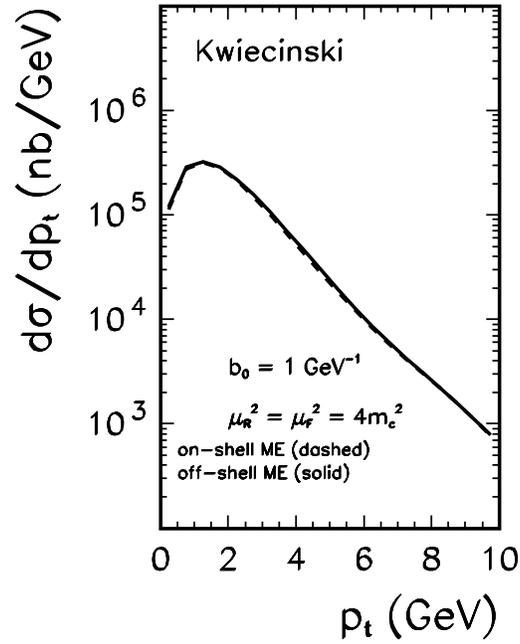}
\caption[*]{
Inclusive $d\sigma/d p_t$ for charm/anticharm production at W = 1.96 TeV
for the Kwieci\'nski UGDF for on-shell (dashed) and off-shell (solid)
matrix element.
In this calculation $b_0$ = 1.0 GeV$^{-1}$, $\mu_R^2$ = $\mu_F^2$ = 4 $m_c^2$.
\label{fig:dsig_dpt_kwiec_onshellme_offshellme}
}
\end{center}
\end{figure}

The off-shell matrix elements necessary for the $k_t$-factorization
approach were calculated only for some selected reactions. 
According to our knowledge it was never shown quantitatively
in the literature what is actual difference if the off-shell matrix
element is replaced by its on-shell counterpart.
In Fig.\ref{fig:dsig_dpt_kwiec_onshellme_offshellme} we present results
for the Kwieci\'nski UGDF for both on-shell and off-shell matrix
elements. For this observable the use of the off-shell matrix element
causes only a small enhancement compared to the use of
the on-shell matrix element.

\begin{figure}[!thb] % Figure 5
\begin{center}
\includegraphics[width=6.0cm]{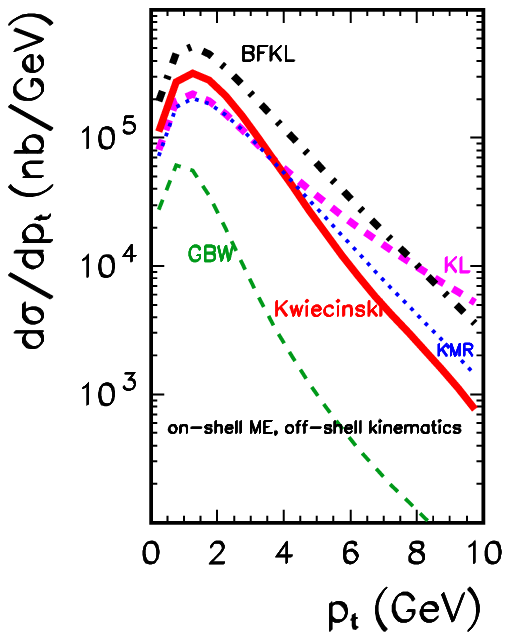}
\includegraphics[width=6.0cm]{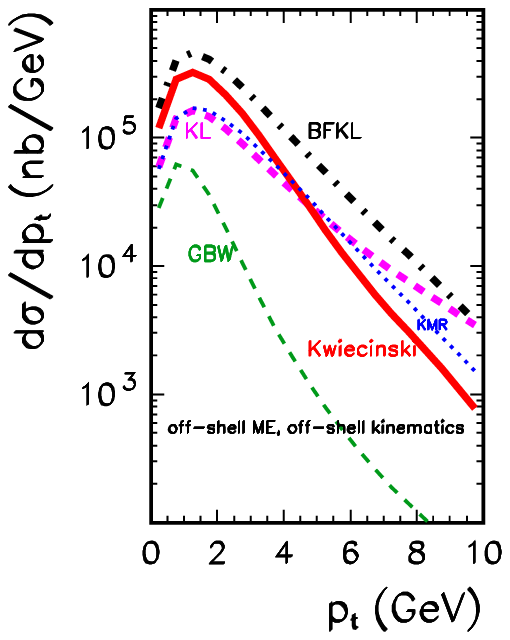}
\caption[*]{
Inclusive $d\sigma/d p_t$ for charm/anticharm production at W = 1.96 TeV
for different UGDFs. The meaning of the curves is as follows:
solid - Kwieci\'nski, thick dashed - KL, thin dashed - GBW,
dash-dotted - BFKL, dotted - KMR.
\label{fig:dsig_dpt_ugdf}
}
\end{center}
\end{figure}

In Fig.\ref{fig:dsig_dpt_ugdf} we collected results obtained with
different unintegrated gluon distributions from the literature.
In this case consequently the off-shell matrix element and
off-shell kinematics were used.
The GBW gluon distribution leads to much smaller cross section.
The KL gluon distribution produces the hardest $p_t$ spectrum.
Rather different slopes in transverse momentum of $c$ (or $\bar c$)
are obtained for different UGDFs.
This differences survive after convoluting the inclusive quark/antiquark
spectra with fragmentation functions.
Thus, in principle, precise distribution in transverse momentum
of charmed mesons should be useful to select a ``correct'' model
of UGDF. A detailed comparison with the experimental data
requires, however, a detailed knoweledge of fragmentation functions.

%-----------

The inclusive spectra are not the best observables to test
UGDF \cite{LS04}.
Let us come now to correlations between charm quark and
charm antiquark.

\begin{figure}[!thb] % Figure 6
\begin{center}
\includegraphics[width=6cm]{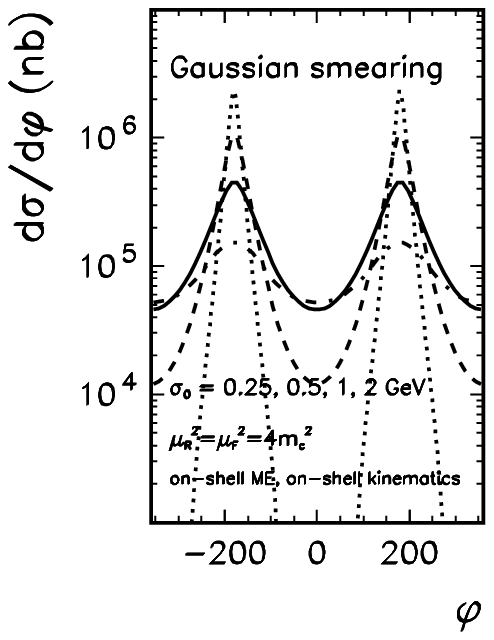}
\includegraphics[width=6cm]{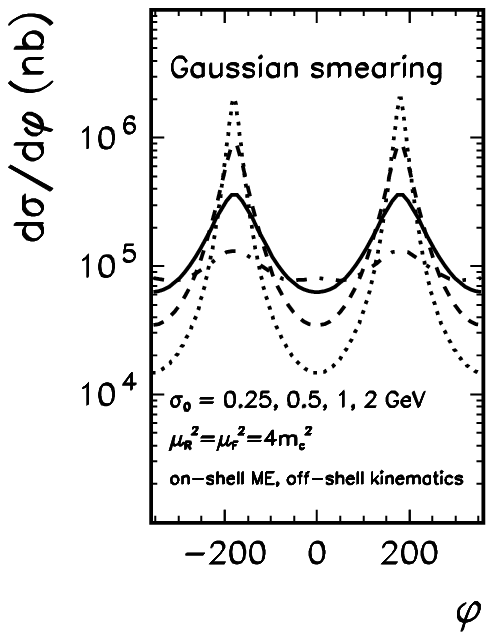}
\caption[*]{
Azimuthal angle correlations for different values of the Gaussian
parameter $\sigma_0$ = 0.25, 0.50, 1.0, 2.0 GeV for on-shell (panel a)
and off-shell (panel b) kinematics.
Here 0 $< p_{1,t}, p_{2,t} <$ 10 GeV.
\label{fig:dsig_dphi_gauss}
}
\end{center}
\end{figure}

\begin{figure}[!thb] % Figure 7
\begin{center}
\includegraphics[width=6.0cm]{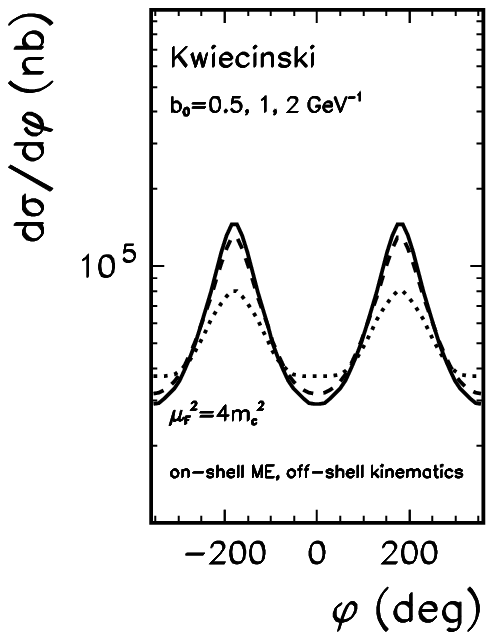}
\includegraphics[width=6.0cm]{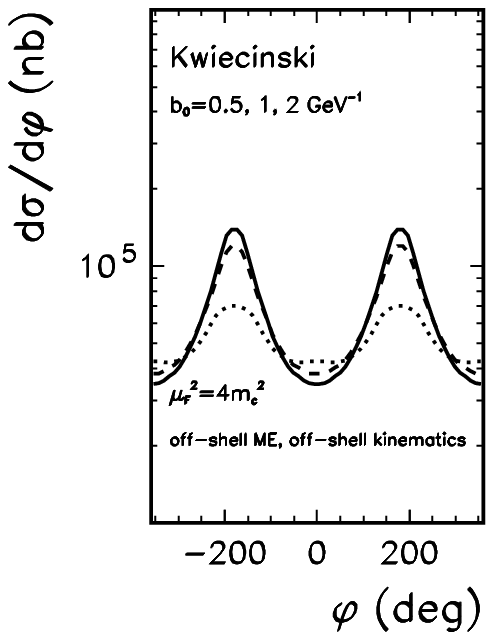}
\caption[*]{
Azimuthal angle correlations for the Kwieci\'nski
UGDF for different values of the parameter $b_0$ = 0.5, 1.0, 2.0 
GeV$^{-1}$. In this calculation $\mu_R^2$ = $\mu_F^2$ = 4 $m_c^2$
was taken.
\label{fig:dsig_dphi_kwiec_b0}
}
\end{center}
\end{figure}

The azimuthal angle correlation is the most popular
observable in this context.
There is a trivial relation between the relative azimuthal angle
distribution and the average transverse momenta of gluons.
This is illustrated in Fig.\ref{fig:dsig_dphi_gauss}
for the simple Gaussian smearing with different values of
the parameter $\sigma_0$ \footnote{The range in azimuthal angle is
somewhat artificially increased to (-360$^0$,360$^0$) to better
visualize the details of the distributions. The normalization
is such that $\int_{-2\pi}^{2\pi} d\sigma/d\phi \; d\phi = \sigma$.} .
In general, the larger $\sigma_0$ the more decorrelation can be
observed.
In the literature both on-shell (see e.g.\cite{AM04})
and off-shell (see e.g.\cite{CCH91,CE91}) kinematics is used.
Our calculation shows that there is almost no difference whether
on-shell or off-shell kinematics is applied.
In Fig.\ref{fig:dsig_dphi_kwiec_b0} we show the angular correlations
for the Kwieci\'nski UGDF for different values of the $b_0$ parameter.
The effect here is smaller than for the simple Gaussian smearing.
This is due to the fact that here the perturbative broadening
is included explicitly and the parameter $b_0$ takes into account
only nonperturbative part. The use of on-shell matrix elements
instead of off-shell ones leads to slightly stronger
back-to-back correlations.

\begin{figure}[!thb] % Figure 8
\begin{center}
\includegraphics[width=6.0cm]{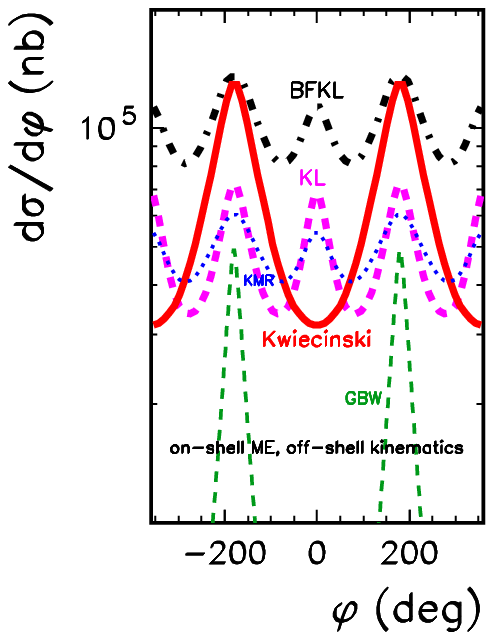}
\includegraphics[width=6.0cm]{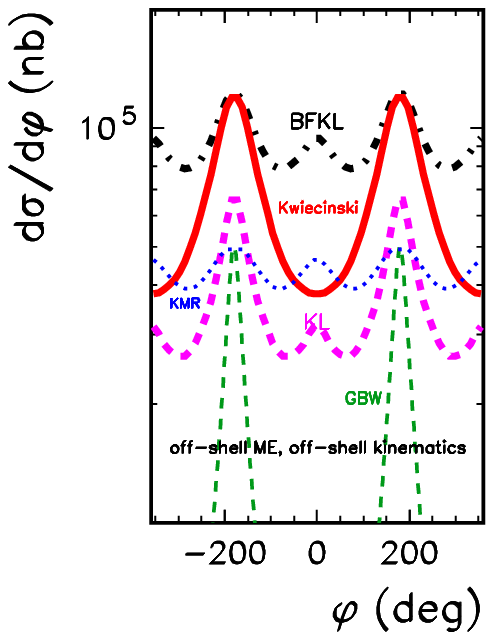}
\caption[*]{
Azimuthal angle correlations for different UGDF in the literature:
Kwieci\'nski (solid), BFKL (dash-dotted), GBW (thin dashed), KL
(thick dashed) and KMR (dotted).
\label{fig:dsig_dphi_ugdf}
}
\end{center}
\end{figure}

In Fig.\ref{fig:dsig_dphi_ugdf} we compare results for different
unintegrated gluon distribution from the literature.
Quite different results are obtained for different UGDF.
The nonperturbative GBW glue leads to strong azimuthal correlations
between $c$ and $\bar c$. In contrast, BFKL dynamics leads to strong
decorrelations of azimuthal angles of charm and anticharm quarks.
The saturation-idea inspired KL distribution, as well as BFKL and
KMR distributions lead to an local
enhancement for $\phi_{c \bar c} \approx$ 0 which is probably due
to gluon splitting s-channel subprocess.
In the last case there is a sizeable difference between the result
obtained with on-shell (left panel) and off-shell (right panel)
matrix elements. Gluon distribution obtained by the KMR method
and the KL and BFKL distributions generate very similar shapes
in azimuthal angle. The gluon distribution obtained by the solution
of the Kwieci\'nski equation gives somewhat narrower distribution
around $\phi = \pm \pi$.

\begin{figure}[!thb] % Figure 9
\begin{center}
\includegraphics[width=6cm]{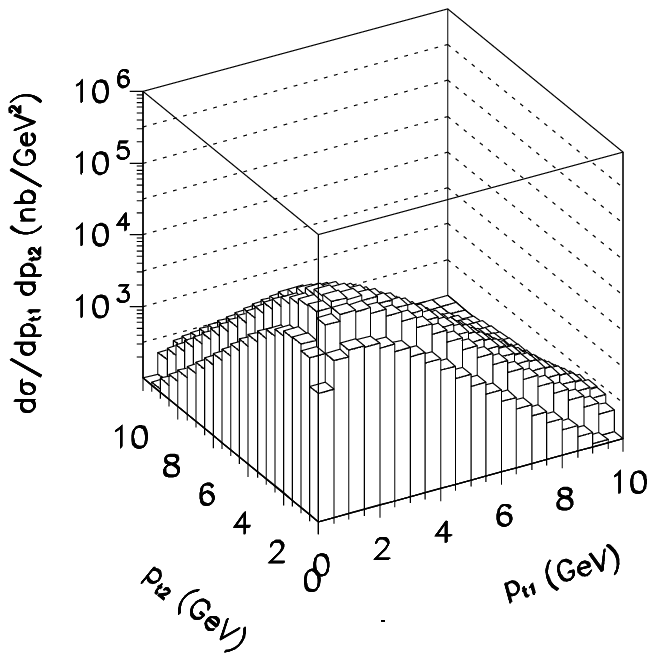}
\includegraphics[width=6cm]{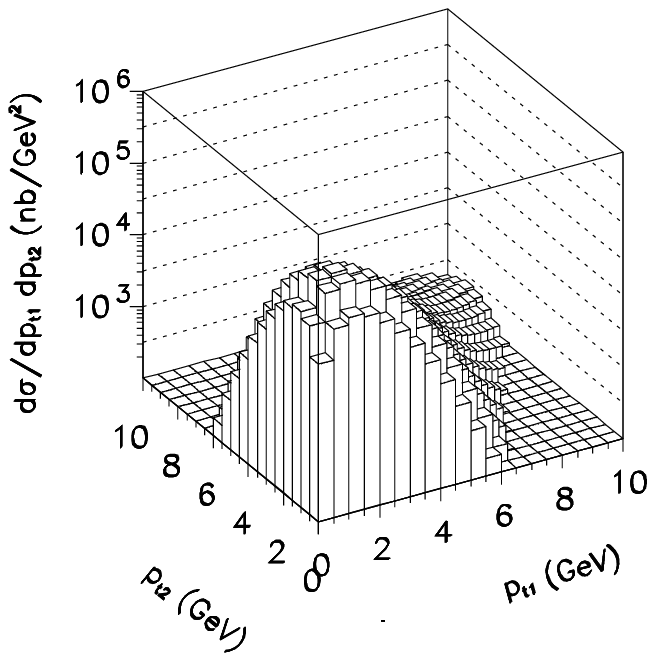}
\includegraphics[width=6cm]{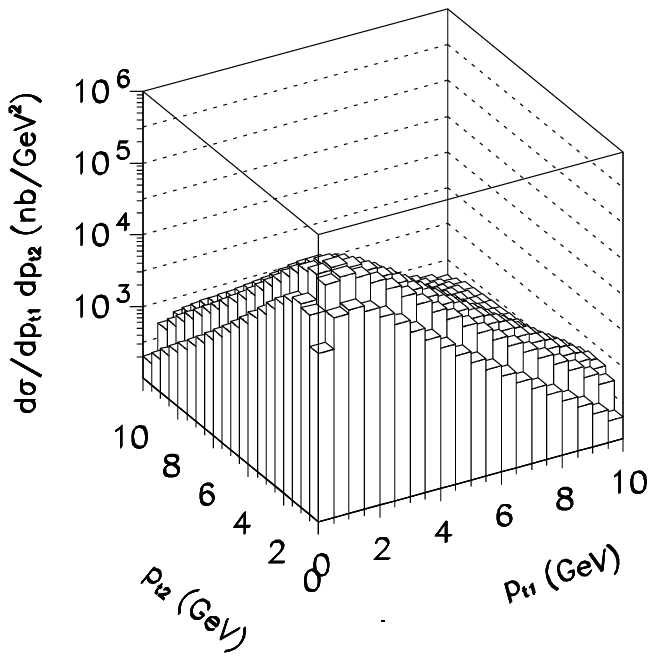}
\includegraphics[width=6cm]{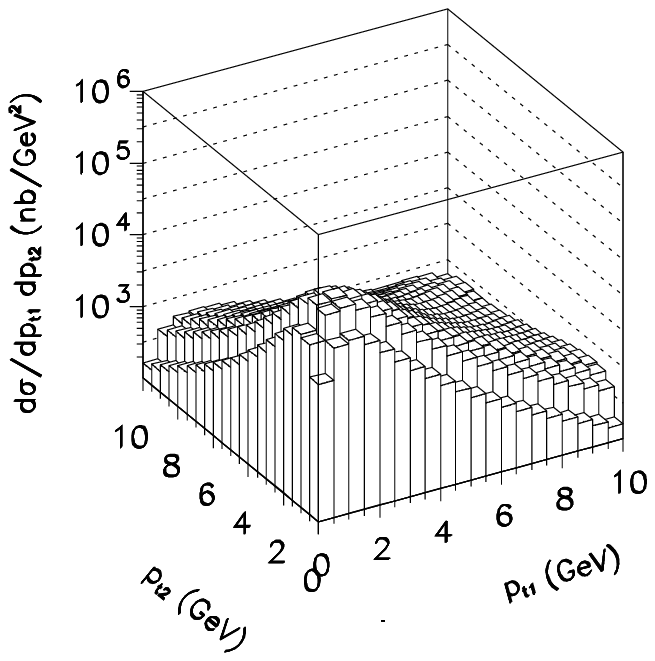}
\caption[*]{
Two-dimensional distributions in $p_{1,t}$ of charm quark and $p_{2,t}$ of
charm antiquark for different UGDF in the literature:
(a) KMR ($\mu_F^2 = 4 m_c^2$),
(b) Kwieci\'nski ($b_0$ = 1.0 GeV$^{-1}$, $\mu_F^2 = 4 m_c^2$),
(c) BFKL,
(d) KL.
\label{fig:maps}
}
\end{center}
\end{figure}

The initial state gluon transverse momentum is also responsible
for the decorrelation of the charm quark, charm antiquark transverse
momenta. In the case of absence of any gluon transverse momenta
in leading-order collinear approximation $p_{1,t} = p_{2,t}$.
The effect of transverse momenta is shown in
Fig.\ref{fig:maps} for
different unintegrated gluon distributions from the literature.
Although the main strength is concentrated along the diagonal
$p_{1,t} = p_{2,t}$, there is a smearing over whole
two-dimensional space of $p_{1,t} \times p_{2,t}$.
As for the case of azimuthal correlations,
different pattern can be observed for different UGDF.

\begin{figure}[!thb] % Figure 10
\begin{center}
\includegraphics[width=6cm]{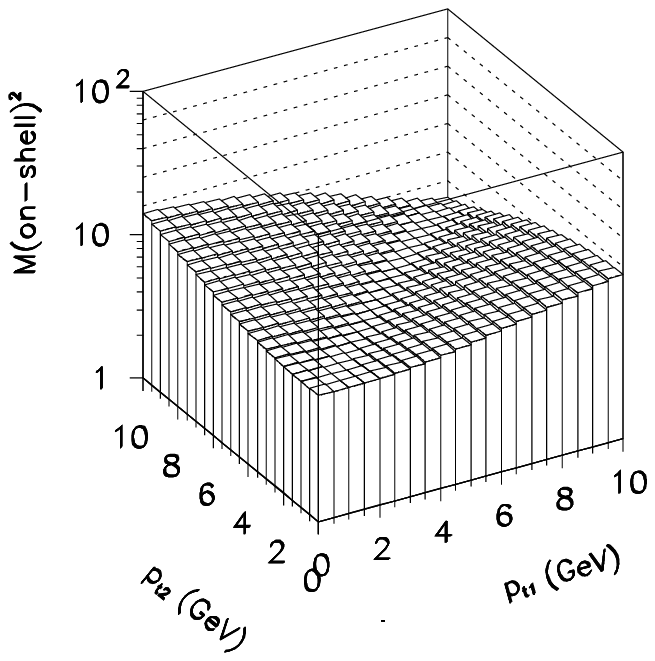}
\includegraphics[width=6cm]{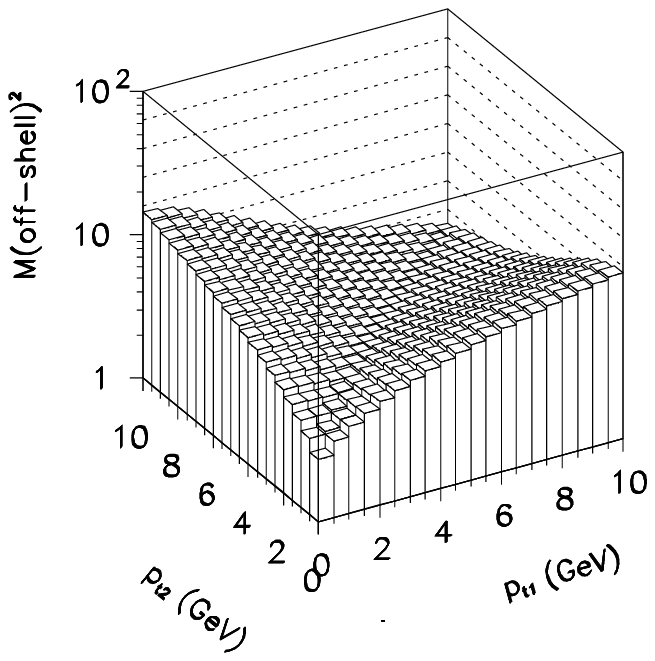}
\caption[*]{
Average matrix element squared as a function of
$p_{1,t}$ of charm quark and $p_{2,t}$ of
charm antiquark for on-shell (left panel) and off-shell (right panel)
matrix element.
In this calculation BFKL UGDF was used.
\label{fig:matrix2}
}
\end{center}
\end{figure}

\begin{figure}[!thb] % Figure 11
\begin{center}
\includegraphics[width=12cm]{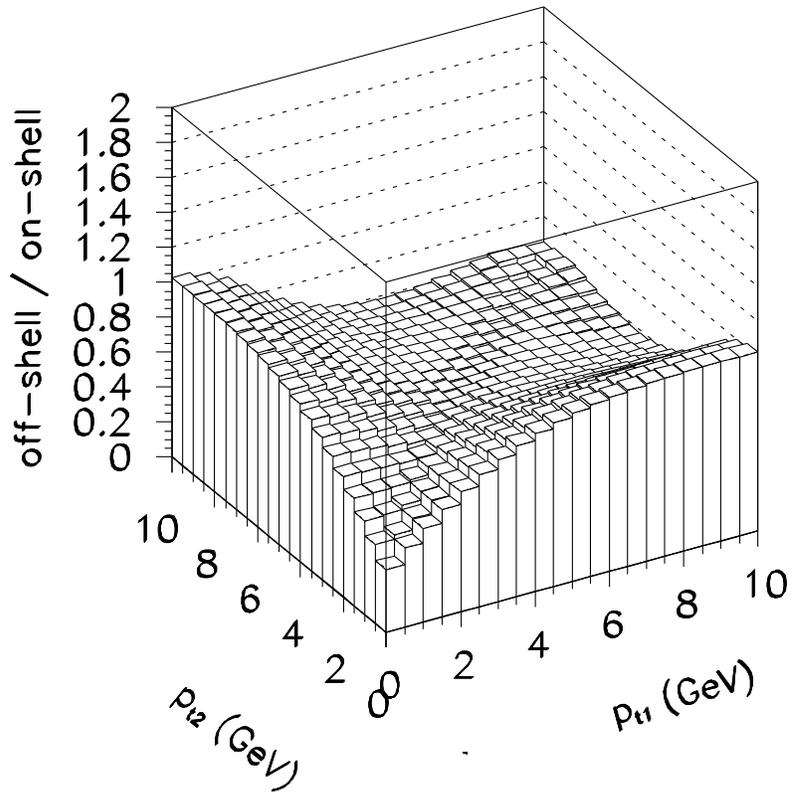}
\caption[*]{
The ratio of the maps (from the previous figure)
with off-shell and on-shell matrix elements.
\label{fig:off-shell-to-on-shell}
}
\end{center}
\end{figure}

In general, the shape of the two-dimensional map
is governed by the UGDF used and the matrix element squared.
It is of interest to unfold these two effects.
In Fig.\ref{fig:matrix2} we show an average matrix element squared
(obtained with the help of BFKL UGDF)
in the $p_{1,t} \times p_{2,t}$ space. A rather weak dependence
can be observed which means that UGDF is the crucial element
responsible for the variation of the cross section
in the $p_{1,t} \times p_{2,t}$ space.
We compare the on-shell (left panel) and off-shell (right panel)
matrix elements squared averaged with UGDF over
the phase space. In contrast to the more inclusive cases
considered above, here the difference is quite sizeable.
In Fig.\ref{fig:off-shell-to-on-shell} we present the corresponding
ratio of both two-dimensional functions.

\begin{figure}[!thb] % Figure 12
\begin{center}
\includegraphics[width=12cm]{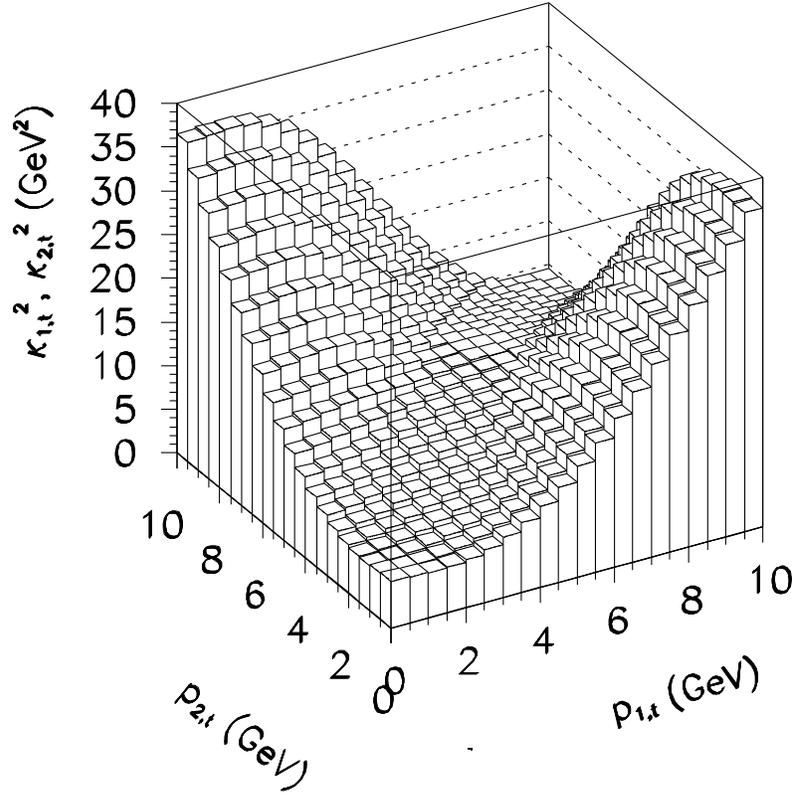}
\caption[*]{
Average transverse momenta of initial gluons 
$\kappa_{1,t}^2$ or $\kappa_{2,t}^2$ as a function
of $p_{1,t}$ and $p_{2,t}$. In this calculation
Kimber-Martin-Ryskin UGDF was used.
\label{fig:KMR_average_kappas}
}
\end{center}
\end{figure}

What is the origin of populating the asymmetric final state
configurations?
The regions $p_{1,t} \gg p_{2,t}$ and $p_{1,t} \ll p_{2,t}$
are populated by asymmetric initial configurations
$\kappa_{1,t}^2 \gg \kappa_{2,t}^2$ or $\kappa_{1,t}^2 \ll
\kappa_{2,t}^2$.
This is demonstrated in Fig.\ref{fig:KMR_average_kappas}
where we show average value of $\kappa_{1,t}^2$ and $\kappa_{2,t}^2$
obtained for example with KMR gluon distributions.
The smallest values of $\kappa_{1,t}^2$ and $\kappa_{2,t}^2$
are obtained along the diagonal ($p_{1,t} = p_{2,t}$).
This means that along diagonal the soft nonperturbative part of UGDF
is tested. The larger distance from the diagonal, the larger
$\kappa_{1,t}^2$ and $\kappa_{2,t}^2$ are sampled.
The KMR is known to have long tails in gluon transverse momenta.
This in conjuction with Fig.\ref{fig:KMR_average_kappas} implies
relatively large cross section for $p_{1,t} \gg p_{2,t}$
or $p_{1,t} \ll p_{2,t}$.

\begin{figure}[!thb] % Figure 13
\begin{center}
\includegraphics[width=6cm]{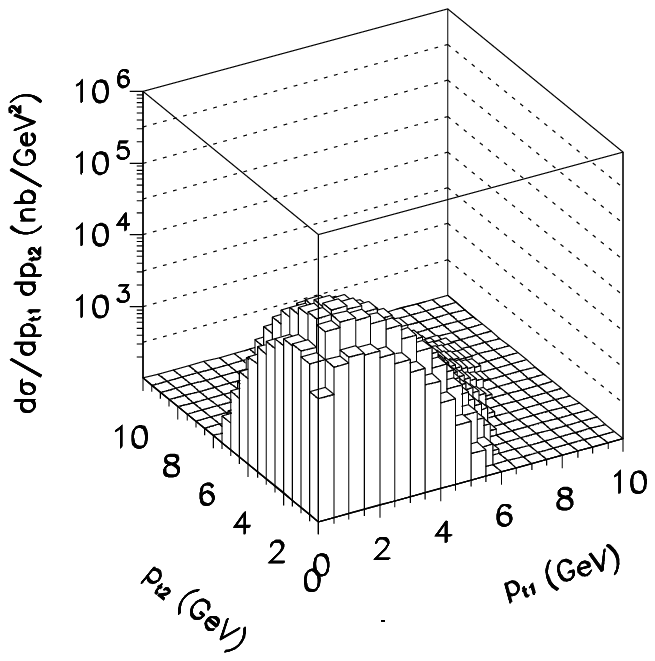}
\includegraphics[width=6cm]{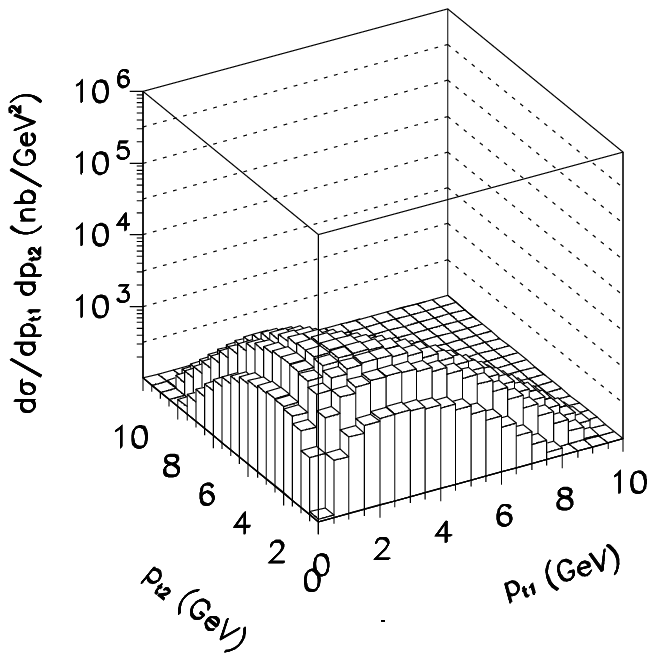}
\includegraphics[width=6cm]{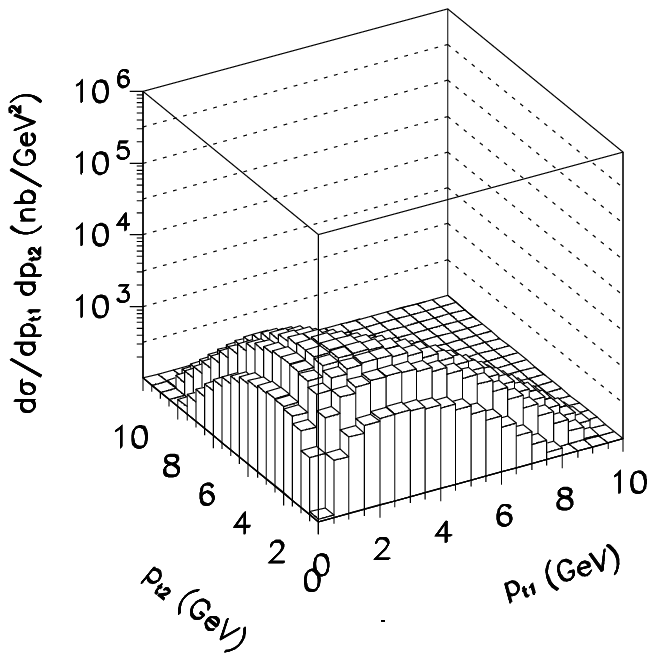}
\includegraphics[width=6cm]{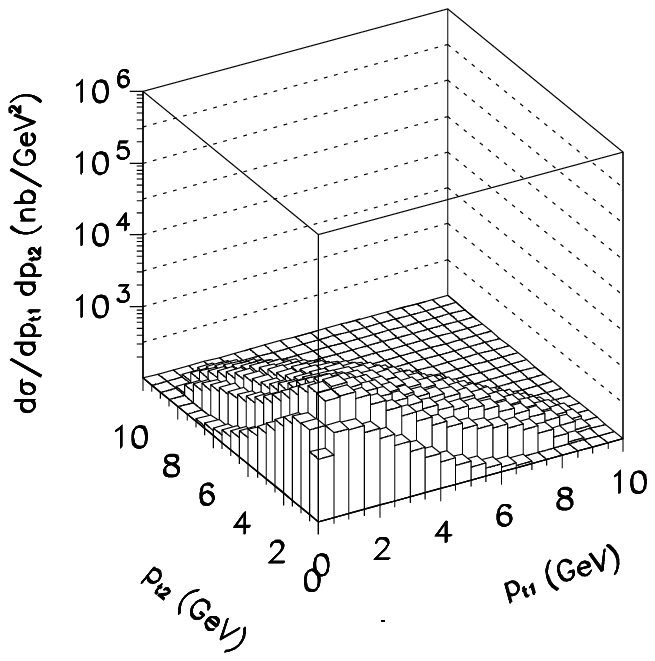}
\caption[*]{
Two-dimensional distributions in $p_{1,t}$ of charm quark and $p_{2,t}$
of charm antiquark for the KMR UGDF and for different regions
of gluon transverse momenta specified in the text.
\label{fig:KMR_cuts_on_kappas}
}
\end{center}
\end{figure}

In Fig.\ref{fig:KMR_cuts_on_kappas} we show two-dimensional maps
in $p_{1,t}$ and $p_{2,t}$ obtained for four different complementary regions
of $\kappa_{1,t}^2$ and $\kappa_{2,t}^2$:\\
(a) $\kappa_{1,t}^2 <$ 10 GeV$^2$, $\kappa_{2,t}^2 <$ 10 GeV$^2$,\\
(b) $\kappa_{1,t}^2 <$ 10 GeV$^2$, $\kappa_{2,t}^2 >$ 10 GeV$^2$,\\
(c) $\kappa_{1,t}^2 >$ 10 GeV$^2$, $\kappa_{2,t}^2 <$ 10 GeV$^2$,\\
(d) $\kappa_{1,t}^2 >$ 10 GeV$^2$, $\kappa_{2,t}^2 >$ 10 GeV$^2$.\\
The asymmetric (in gluon transverse momentum) configurations
contribute a large fraction to the integrated cross section.
In Table 1 we ensambled fractions of different regions of
$\kappa_1^2 \times \kappa_2^2$. For completeness we placed in the table
the integrated cross section for the $c \bar c$ pair production.
The KMR and KL have largest fractions of asymmetric initial state
configurations.
The almost vanishing fraction of asymmetric
configurations for GBW glue is due to a lack of pQCD effects.
A rather schematic KL distribution includes the hard components only in a
simple parametric form. Our purely model discussion here is a bit academic
because initial state configurations are not directly the observables.
Therefore the effect of the large tails in UGDFs can be only
tested indirectly by studying asymmetric configurations in
the final state (charmed mesons, muons) space $p_{1,t} \times p_{2,t}$.

The decorrelation in $p_{1,t}$ and $p_{2,t}$ can be also studied
in the collinear next-to-leading order approach.
It is worth to stress here, however, that collinear approach even in
higher order approximation is unreliable in the region
$p_{1,t} \approx p_{2,t}$.
In order to avoid conceptual and/or numerical problems
asymmetric cuts on jets transverse momentum are usually assumed.
On experimental side this means removing a big portion
of statistics from the correlation studies.
Our approach is free of the mentioned above problems
and can be applied on the whole plane $p_{1,t} \times p_{2,t}$.

\begin{figure}[!thb] % Figure 14
\begin{center}
\includegraphics[width=6cm]{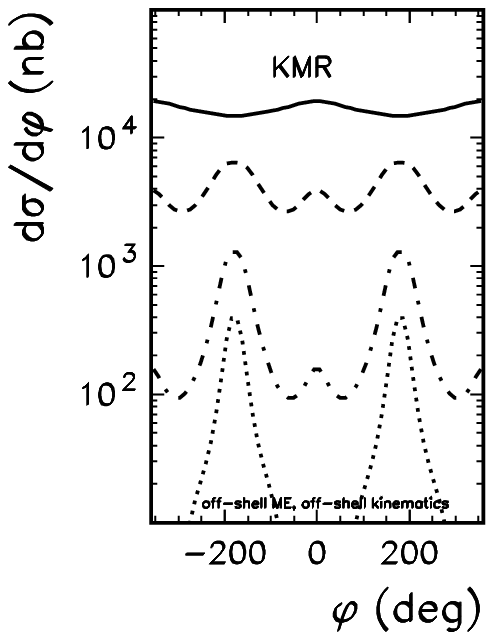}
\includegraphics[width=6cm]{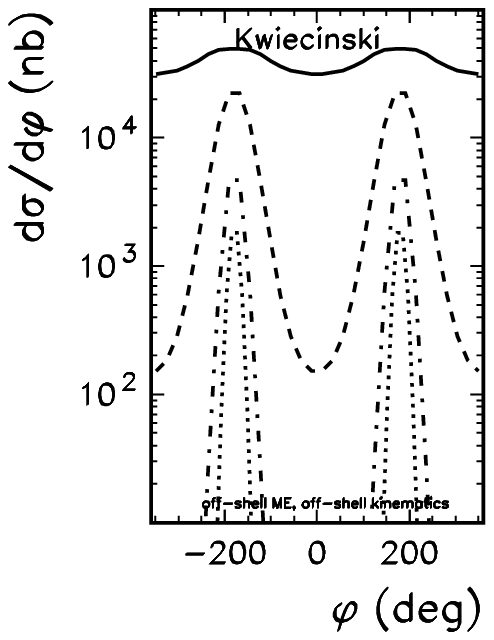}
\includegraphics[width=6cm]{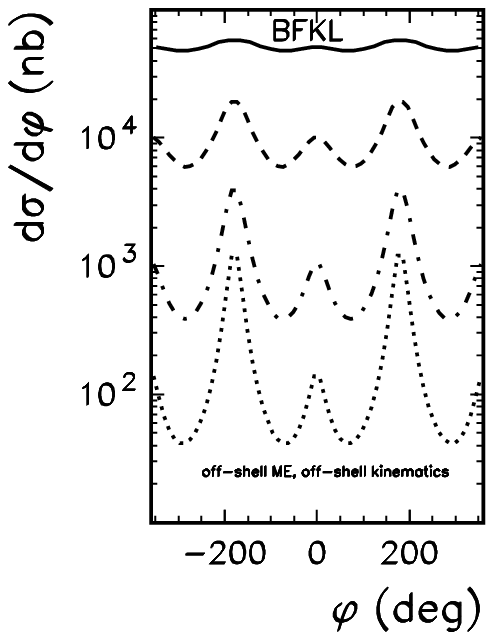}
\includegraphics[width=6cm]{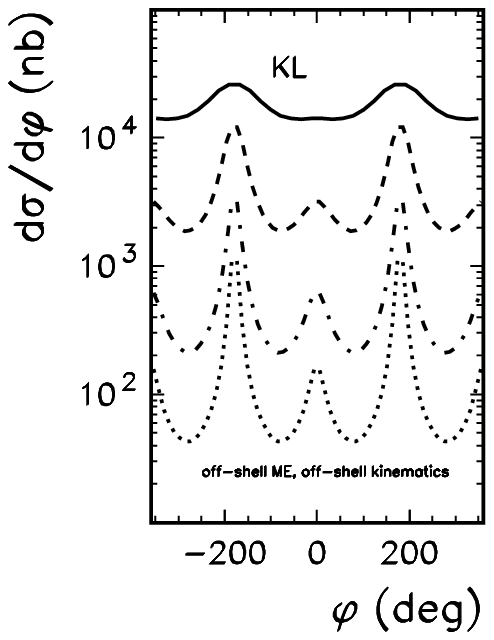}
\caption[*]{
$d\sigma/d\phi_{-}$ for different intervals of charm and anticharm
transverse momenta:\\
0.0 GeV $< p_{1,t},p_{2,t} <$ 2.5 GeV (solid),\\
2.5 GeV $< p_{1,t},p_{2,t} <$ 5.0 GeV (dashed),\\
5.0 GeV $< p_{1,t},p_{2,t} <$ 7.5 GeV (dash-dotted),\\
7.5 GeV $< p_{1,t},p_{2,t} <$ 10.0 GeV (dotted) \\
for different UGDF:
(a) KMR ($\mu^2 = 4 m_c^2$),
(b) Kwieci\'nski ($b_0$ = 0.5 GeV$^{-1}$, $\mu^2 = 4 m_c^2$),
(c) BFKL,
(d) KL.
\label{fig:dsig_dphi_sym}
}
\end{center}
\end{figure}

In principle, there are many more correlation observables possible
having in view quite reach phase space for two particle
production.
Here we shall concentrate on one more observable -- namely
the azimuthal angle correlation for different regions
of quark-antiquark transverse momenta.
In Fig.\ref{fig:dsig_dphi_sym} we show such correlations
but now with extra conditions on quark-antiquark transverse momenta.
Generally, the larger transverse momenta the more back-to-back
correlation is observed. The details depend, however,
on a particular model of UGDF.

\begin{figure}[!thb] % Figure 15
\begin{center}
\includegraphics[width=10.0cm]{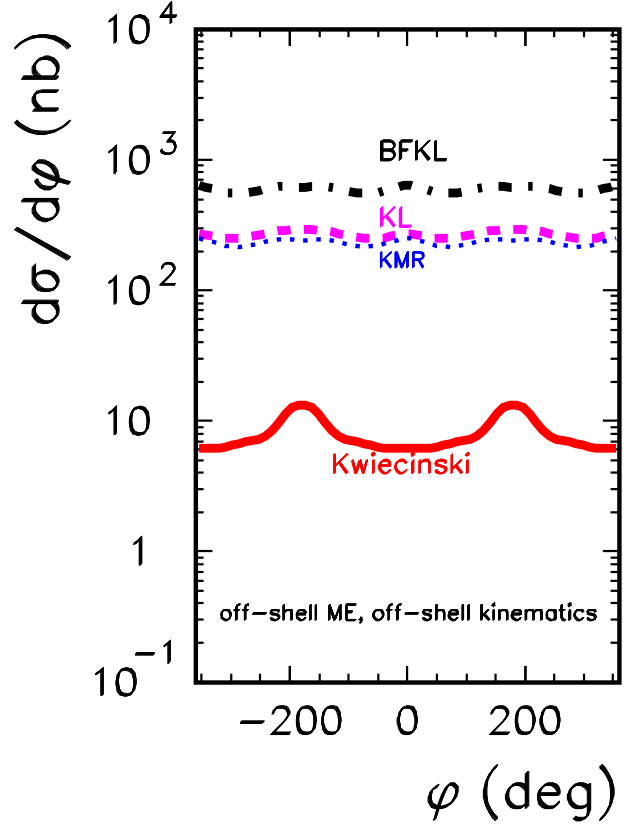}
\caption[*]{
$d\sigma/d\phi_{-}$ for \\
0.0 GeV $< p_{1,t} <$ 2.5 GeV and 7.5 GeV $< p_{2,t} <$ 10. GeV or\\
7.5 GeV $< p_{1,t} <$ 10. GeV and 0.0 GeV $< p_{2,t} <$ 2.5 GeV \\
for different UGDF:
KMR ($\mu_F^2 = 4 m_c^2$) -- dotted,
Kwieci\'nski ($b_0$ = 0.5 GeV$^{-1}$, $\mu_F^2 = 4 m_c^2$) -- solid,
BFKL -- dash-dotted,
and KL -- dashed.
\label{fig:dsig_dphi_asym}
}
\end{center}
\end{figure}

In Fig.\ref{fig:dsig_dphi_asym} we show similar correlations for
very asymmetric values of $p_{1,t}$ and $p_{2,t}$. 
In this domain the cross section obtained with BFKL, KL and KMR UGDFs
is much larger than the cross section obtained with the Kwieci\'nski
UGDF. While the first three UGDFs lead to almost
complete decorrelation of $c$ and $\bar c$, for the last UGDF
we observe clear back-to-back correlations. Furthermore
the absolute cross section in the latter cases is much smaller.
This distinct difference causes that this observable is very promissing
in testing the QCD dynamics.

\begin{figure}[!thb] % Figure 16
\begin{center}
\includegraphics[width=6.0cm]{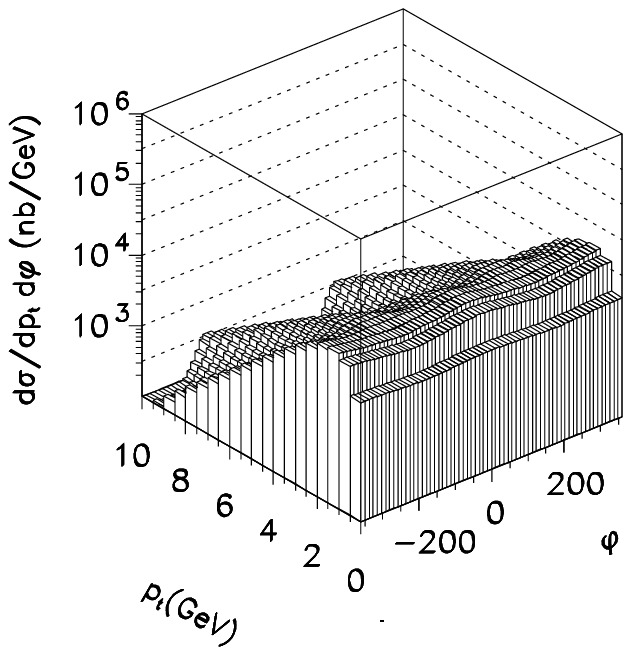}
\includegraphics[width=6.0cm]{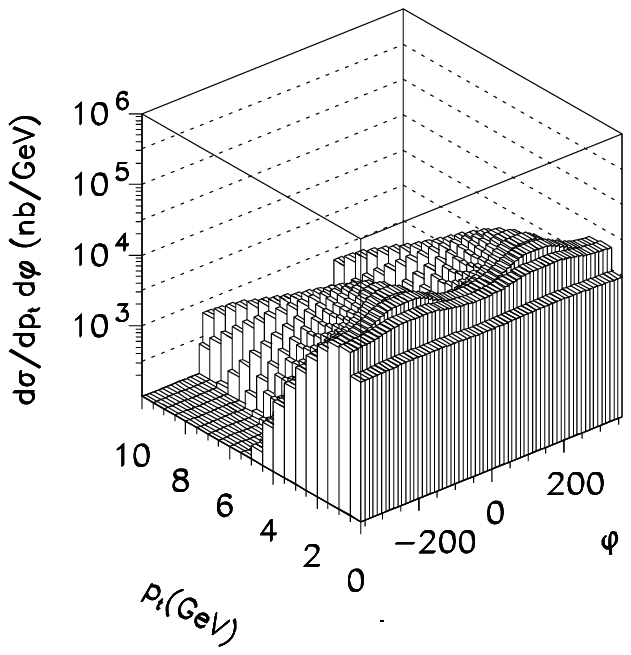}
\includegraphics[width=6.0cm]{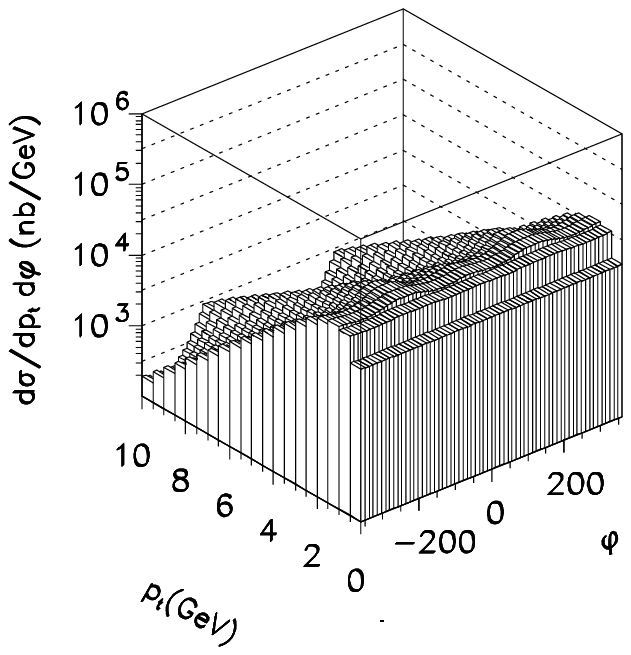}
\includegraphics[width=6.0cm]{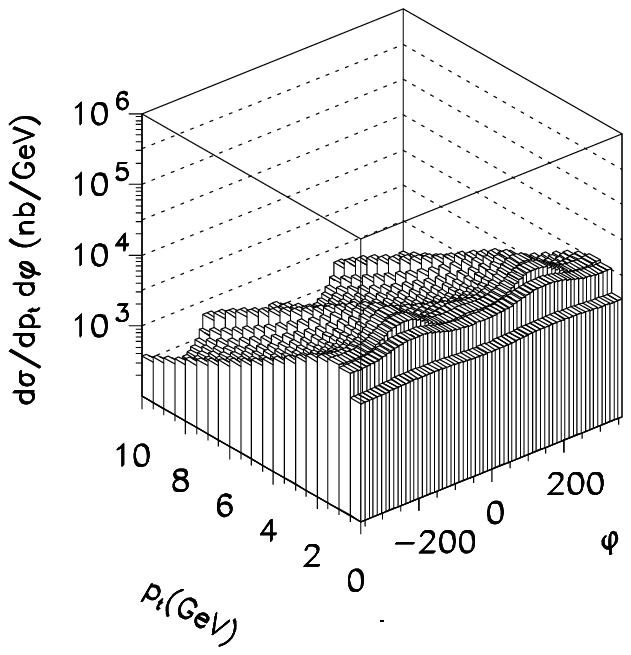}
\caption[*]{
$d\sigma/d p_{t} d \phi_{-}$ for different UGDF:
(a) KMR ($\mu_F^2 = 4 m_c^2$),
(b) Kwieci\'nski ($b_0$ = 0.5 GeV$^{-1}$, $\mu_F^2 = 4 m_c^2$),
(c) BFKL,
(d) KL.
\label{fig:dsig_dptdphi}
}
\end{center}
\end{figure}

In order to shed more light on the relation between
the $c$/$\bar c$ transverse momentum and azimuthal correlations
in Fig.\ref{fig:dsig_dptdphi} we show the effect of transverse momentum
of $c$ or $\bar c$ and relative azimuthal angle
on a two-dimensional map $(p_t,\phi_{-})$, where $p_t$ is transverse
momentum of either quark or antiquark.
The same effect as in Fig.\ref{fig:dsig_dphi_sym} is shown now
in a continuous way.
There is also a technical lesson from the inspection of
Fig.\ref{fig:dsig_dptdphi}.
The figure shows that one needs somewhat more refine integration
in $\phi_{-}$ for larger values of $p_{1,t}$ and $p_{2,t}$.

In the present exploratory calculation we have studied correlations
between charm quark and charm antiquark. 
In practice one measures either muons or charmed mesons.
It seems that a study of $p_t(D^{*+}) \times p_t(D^{*-})$
or $p_t(\mu^+) \times p_t(\mu^-)$ or mixed
$p_t(D^{*+}) \times p_t(\mu^-)$ or
$p_t(\mu^+) \times p_t(D^{*-})$  correlations
would be optimal to test UGDFs. This will be a subject of our next studies.
It is not clear to us if such measurements are feasible with
the present Tevatron apparatus.

%--------------------
\section{Conclusions}
%--------------------

We have compared quantitatively different methods to include
gluon transverse momenta and their effect on inclusive spectra
as well as on $c \bar c$ correlations. Different UGDFs from
the literature were used.

The gluon-gluon luminosity factor is more strongly varying factor over the
phase space than the matrix element squared. Consequently,
the shape of inclusive spectra only weakly depends on whether on-shell
or off-shell matrix element is used. In contrast there is
a sizeable effect for more exclusive variables, dependent e.g. on
the position in the $p_{1,t} \times p_{2,t}$ space.

We have shown that the analysis of azimuthal correlations
and decorrelations on the
$p_{1,t}$(c transverse momentum) x $p_{2,t}$($\bar c$ transverse
momentum) plane as well as the combination of both
are very promissing to verify UGDF.
We have focussed on the region of $p_{1,t} \gg p_{2,t}$ and
$p_{1,t} \ll p_{2,t}$. There max$\{\kappa_{1,t},\kappa_{2,t}\} \sim$
max$\{p_{1,t},p_{2,t}\}$ and off-shell matrix element must be used.
In this region the shape of azimuthal correlation function
depends on both UGDF and matrix element.

Although we have calculated correlation observables for
charm quarks and antiquarks it is obvious that the main
effects should be very similar for charmed mesons and/or muons.
The correlations between $D^{*+}(c \bar d)$ and $D^{*-}(\bar c d)$
or $\mu^+$ and $\mu^-$ are the most promissing in this respect as they
reflect relatively well the $c \bar c$ final state.
Estimating the rates of events with one $D$ meson with small transverse
momentum and second $D$ meson with large transverse momentum
would be a useful test of the QCD dynamics.
The analysis of meson correlations will be a subject of our
next analysis.

\vskip 0.5cm

{\bf Acknowledgements}
We are indebted to Sergey Baranov for an interesting discussion
on heavy quark - heavy antiquark correlations.
This work was partially supported by the grant of the Polish
Ministry of Scientific Research and Information Technology
number 1 P03B 028 28.

%======================================================================

\newpage

{\bf TABLES}

%-----------------------------------------------------------------------

\begin{table}

\caption{Fractions in \% of contributions from different regions of
 $\kappa_1^2 \times \kappa_2^2$
to the total cross section for different UGDFs from the literature
for W = 1.96 TeV, with off-shell kinematics and off-shell matrix elements.
In the case of Kwieci\'nski UGDF: $b_0$ = 1 GeV$^{-1}$, $\mu_F^2$ = 4
$m_c^2$, in the case of KMR UGDF: $\mu_F^2$ = 4 $m_c^2$.
Below for brevity $\xi^2$ = 10 GeV$^2$ is assumed.}

\begin{center}

\begin{tabular}{|c|c|c|c|c|c|}
\hline 
UGDF & $\sigma_{c \bar c}^{tot}$(mb) & 
$\kappa_1^2 < \xi^2 , \kappa_2^2 < \xi^2 $ &
$\kappa_1^2 < \xi^2 , \kappa_2^2 > \xi^2 $ &
$\kappa_1^2 > \xi^2 , \kappa_2^2 < \xi^2 $ &
$\kappa_1^2 > \xi^2 , \kappa_2^2 > \xi^2 $ \\
\hline
Kwieci\'nski & 0.85   & 91.87  & 3.99   & 3.99   & 0.15   \\
KMR          & 0.55   & 52.43  & 19.83  & 19.83  & 7.91   \\
BFKL         & 1.44   & 59.65  & 17.26  & 17.26  & 5.83  \\
GBW          & 0.12   & 100.0  & 0.001  & 0.001  & 0.0   \\
KL           & 0.51   & 65.49  & 15.37  & 15.37  & 3.76  \\
\hline
\end{tabular}

\end{center}

\end{table}

\newpage

{\bf FIGURES}

%--------------------------------------------------------------------

%--------------------------------------------------------------------

%--------------------------------------------------------------------

%--------------------------------------------------------------------

%--------------------------------------------------------------------

%--------------------------------------------------------------------

%-----------------------------------------------------------------

%-----------------------------------------------------------------

%--------------------------------------------------------------------

%---------------------------------------------------------------------

%---------------------------------------------------------------------

%---------------------------------------------------------------------

%---------------------------------------------------------------------

%---------------------------------------------------------------------

%---------------------------------------------------------------------

%---------------------------------------------------------------------

%---------------------------------------------------------------------

\end{document}